\documentclass[11pt,twoside]{article}

\usepackage{asp2014}

\aspSuppressVolSlug
\resetcounters

\begin{document}

\title{A Study of the Efficiency of Spatial Indexing Methods Applied to Large Astronomical Databases}

\author{G. Bruce Berriman,$^1$ John C. Good,$^1$ Bernie Shiao,$^2$ and Tom Donaldson $^2$
\affil{$^1$ Caltech/IPAC, Pasadena, CA  91125, USA}  \email{gbb@ipac.caltech.edu}}
\affil{$^2$Space Telescope Science Institute, 3700 San Martin Drive, Baltimore, MD  21218, USA}

\paperauthor{G. Bruce Berriman}{gbb@ipac.caltech.edu}{0000-0001-8388-534X}{Caltech}{IPAC-NEXScI}{Pasadena}{CA}{91125}{USA}
\paperauthor{John C. Good}{jcg@ipac.caltech.edu}{ORCID_Or_Blank}{Caltech}{IPAC-NEXScI}{Pasadena}{CA}{91125}{USA}
\paperauthor{Bernie Shiao}{shiao@stsci.edu}{ORCID_Or_Blank}{Space Telescope Science Institute}{ }{Baltimore}{MD}{21218}{USA}
\paperauthor{Tom Donaldson}{tdonaldson@stsci.edu}{ORCID_Or_Blank}{Space Telescope Science Institute}{ }{Baltimore}{MD}{21218}{USA}

\begin{abstract}
Spatial indexing of astronomical databases generally uses quadrature methods, which partition the sky into cells to create an index (usually a B-tree) written as a database column. We report the results of a study to compare the performance of two common indexing methods, HTM and HEALPix, on Solaris and Windows database servers installed with PostgreSQL, and a Windows Server installed with MS SQL Server. The indexing was applied to the 2MASS All-Sky Catalog and to the Hubble Source Catalog, which approximate the diversity of catalogs common in astronomy. The study used a dedicated software package in ANSI-C for creating database indices and constructing queries, which will be released in winter 2017. On each server, the study compared indexing performance by submitting 1 million queries at each index level with random sky positions and random cone search radius, which was computed on a logarithmic scale between 1 arcsec and 1 degree, and measuring the time to complete the query and write the output. These simulated queries, intended to model realistic use patterns, were run in a uniform way on many combinations of indexing method and indexing depth. The query times in all simulations are strongly I/O-bound and are linear with number of records returned for large numbers of sources. There are, however, considerable differences between simulations, which reveal that hardware I/O throughput is a more important factor in managing the performance of a DBMS than the choice of indexing scheme. The choice of index itself is relatively unimportant: for comparable index levels, the performance is consistent within the scatter of the timings. At small index levels (large cells; e.g. level 4; cell size 3.7 deg), there is large scatter in the timings because of wide variations in the number of sources found in the cells. At larger index levels, performance improves and scatter decreases, but the improvement at level 8 (14 arcmin) and higher is masked to some extent in the timing scatter caused by the range of query sizes. At very high levels (20; 0.0004 arsec), the granularity of the cells becomes so high that a large number of extraneous empty cells begin to degrade performance. Thus, for the use patterns studied here, the database performance is not critically dependent on the exact choices of index or level.
\end{abstract}

\section{Introduction}
Spatial indexing methods used in astronomy are usually based on quadrature methods. These methods partition the sky into cells, use the cell numbers to create an index, usually a binary tree (B-tree), and write the index as a database column. What determines the performance of a database index: the choice of index? Its depth? The choice of hardware?

We have therefore compared the performance of two indexing schemes, the Hierarchical Triangular Mesh (HTM, \citet{2000ASPC..216..141K}) and the Hierarchical Equal Area iso-Latitude Pixelation of the sphere (HEALPix, \citet{2005ApJ...622..759G}), on Solaris, Windows and Red Hat Linux platforms. The indexing was applied to the 2MASS All-Sky Point Source Catalog (470 million records), and the  non-merged Hubble Source Catalog (384 million records), installed in PostgreSQL and SQL Server databases. Table 1 summarizes the platforms and databases used. The study was aimed at comparing the performance of indexing schemes, and therefore, neither caching of records in memory nor clustering (ordering) of data within the database was permitted.

\begin{table}[!ht]
\caption{Summary of Hardware Platforms and Database Management Systems}
\smallskip
\begin{center}
{\small
\begin{tabular}{llll}  
\tableline
\noalign{\smallskip}
Center & Database & OS/Compiler & Server\\
\noalign{\smallskip}
\tableline
\noalign{\smallskip}
 STScI & PostgreSQL  9.5  &  Win 2012 server OS                             & 2 processors @3.46 GHz                                  \\
~         & ~                           &  Cygwin 2.5.2 DLL                & 6 cores. 			 \\
\tableline
\noalign{\smallskip}
STScI & SQL Server 2012 &Win 2012 server OS       & 2 processors @3.46 GHz  \\
~        &         ~                      & Cygwin 2.5.2 DLL & 6 cores \\
\tableline
\noalign{\smallskip}
IPAC    &PostgreSQL  9.3.5 & Solaris 10  &   2 processors @ 2.27 GHz.                                            \\
~          &  ~                            & ~                &  4 cores                                    \\
\noalign{\smallskip}
\tableline
\noalign{\smallskip}
 IPAC    &PostgreSQL  9.4 & Red Hat Linux 4.2.1  &  8 processors  @ 2.66 GHz                                      \\
\noalign{\smallskip}
\tableline
\\
\end{tabular}
}
\end{center}
\end{table}
\noindent 

\section {Methodology}

The methodology was simple. The 2MASS and HSC catalogs were stripped of all fields except source designation, right ascension and declination. Added to each record were a spatial index value for a specified HTM/HEALPix level between 4 and 20, and the x, y, z spherical 3-vector values. The table was ingested into the database, and a B-tree index was created for the index column. For 1 million queries with random sky positions and cone search radii, selected on a log scale between 1 arcsec and 1 degree, the locations and radii were used to create a list of spatials bins that intersect each region, and the sources in these bins were filtered with the 3-vector values to derive the final source count within the input cone. The computations were performed with a dedicated software package written in ANSI-C and including all support libraries.  Release of the code, attached with a BSD 3-clause license, is planned for winter 2017.

\section{What Do The Experiments Show?}

It will be easiest to state the conclusions first, and then refer to the figures. Where comparable, the results are consistent with those of \citet{O07-3_adassxxvi}.

\begin{itemize}
\checklistitemize
\item The total query time is dominated by I/O and is linear with the number of records returned for large numbers of sources.  For each platform, there is a start-up time that is nearly constant for  all experiments.  See Figure \ref{Fig 1}.
\item Depth of tesselation and hardware configuration have greater impact on database performance than the choice of tesselation scheme. See Figure \ref{Fig 1}.
\item The optimum indexing level corresponds roughly to when the average search radius is of the order of the cell size. Performance does not improve as the indexing level becomes deeper. See Figures \ref{Fig 2} and \ref{Fig 3}.  At shallow index depths, performance is slowed down in dense regions because large numbers of sources have to be filtered post-query. The numbers of such extraneous sources decline as the index becomes deeper, until near level 20, the performance slows down as the query must search through an ever larger number of empty cells.
\item The optimum depth of indexing level depends only weakly on the spatial distribution of queries.   See Figure \ref{Fig 4}.
\end{itemize}

\articlefigure[width=.5\textwidth,height=2.15in]{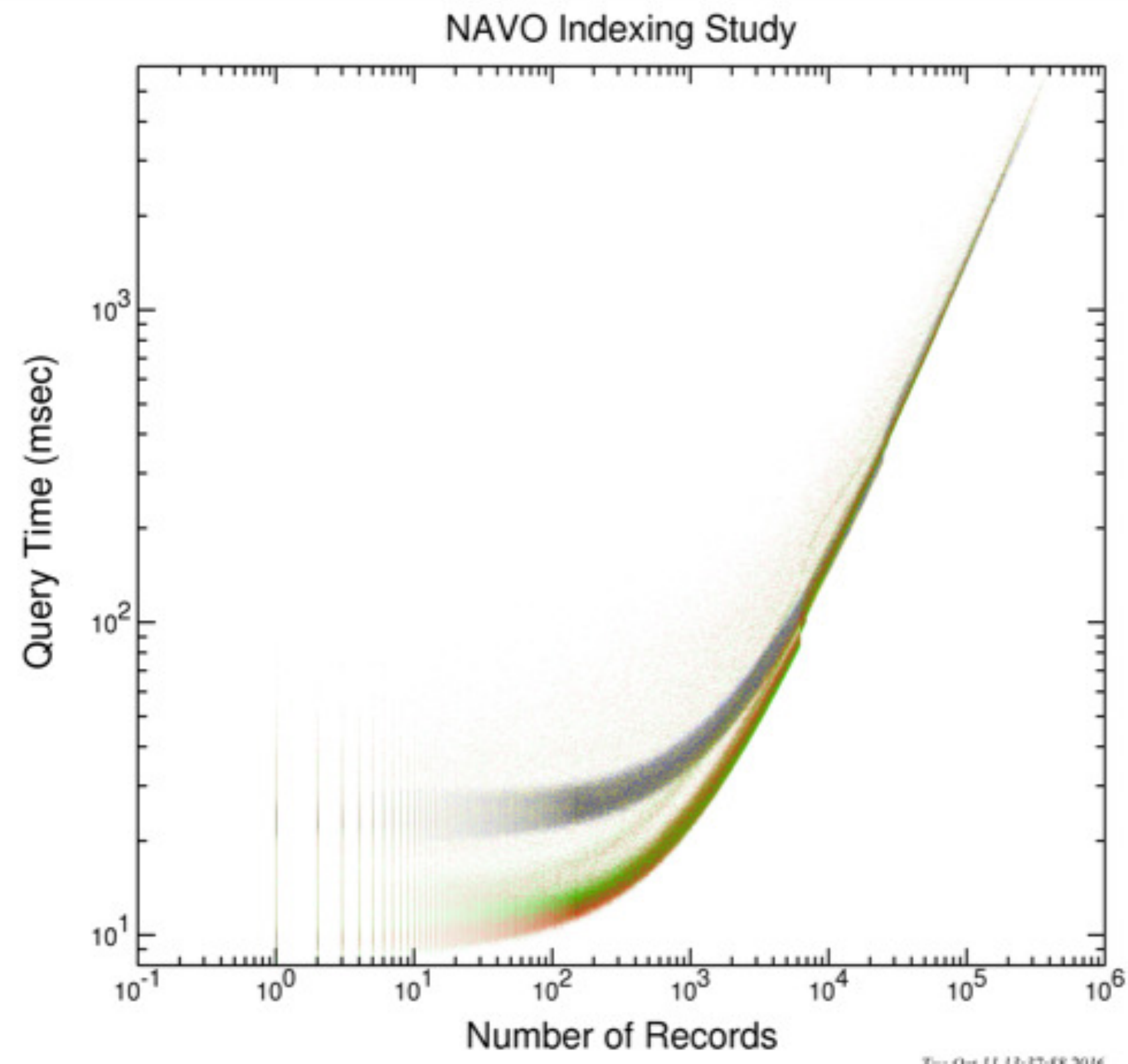}{Fig 1}{Distribution of query times versus number of records returned for the 2MASS catalog for indexing at level 14 in PostgreSQL. Key:  Blue: Windows, HPX; Yellow: Windows, HTM; Red: Solaris, HPX; Green: Solaris, HTM.}

\articlefigurethree {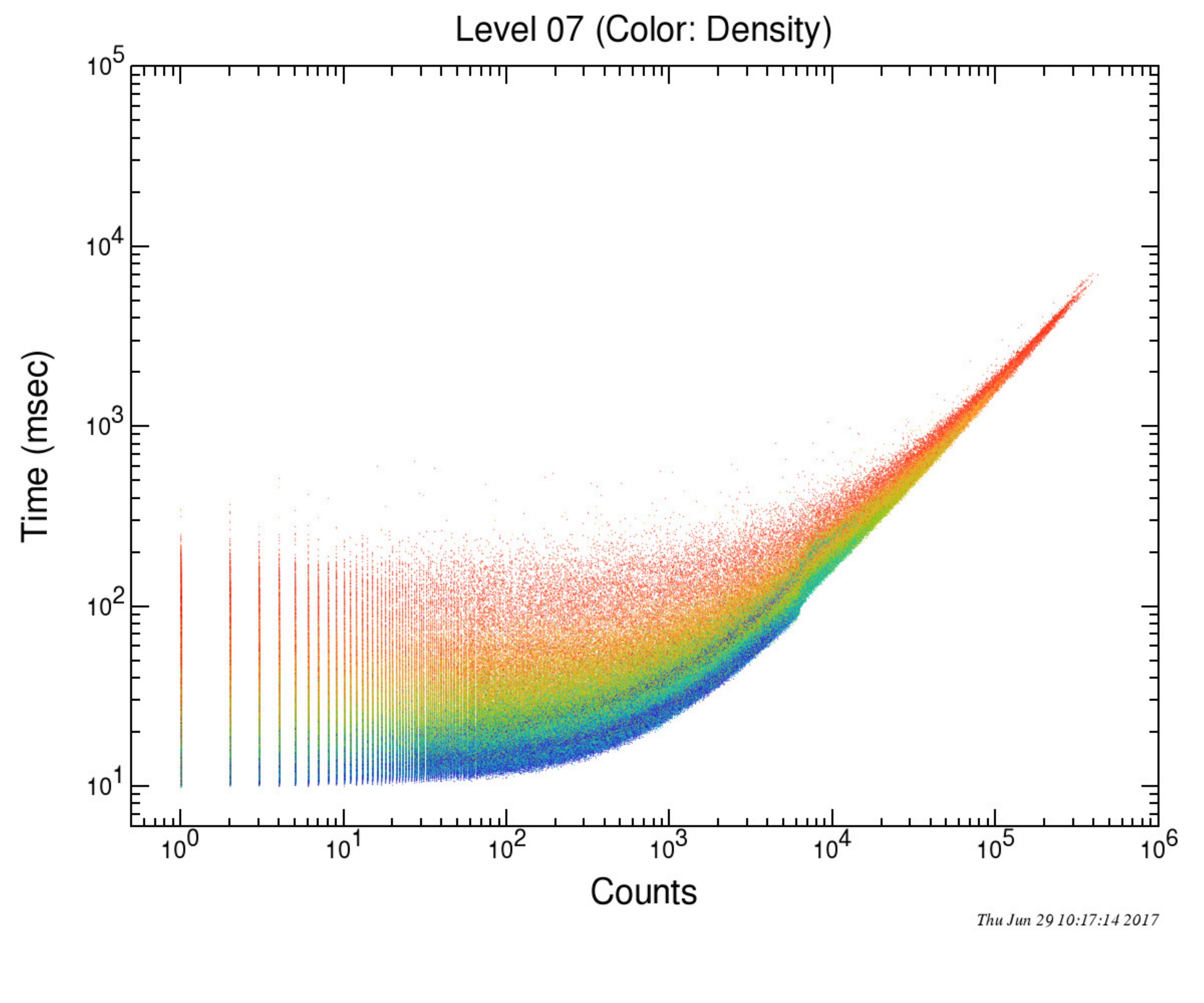}{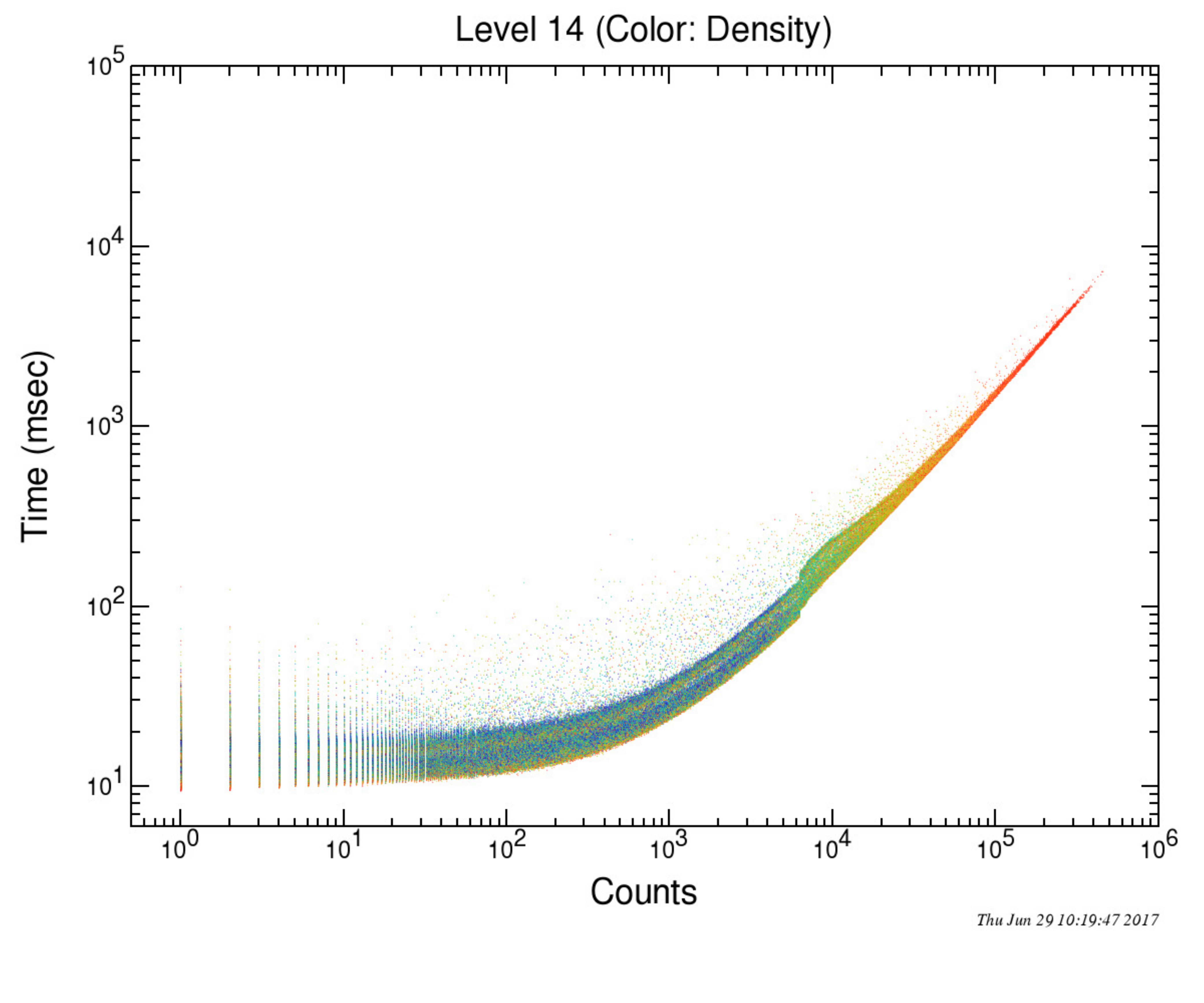}{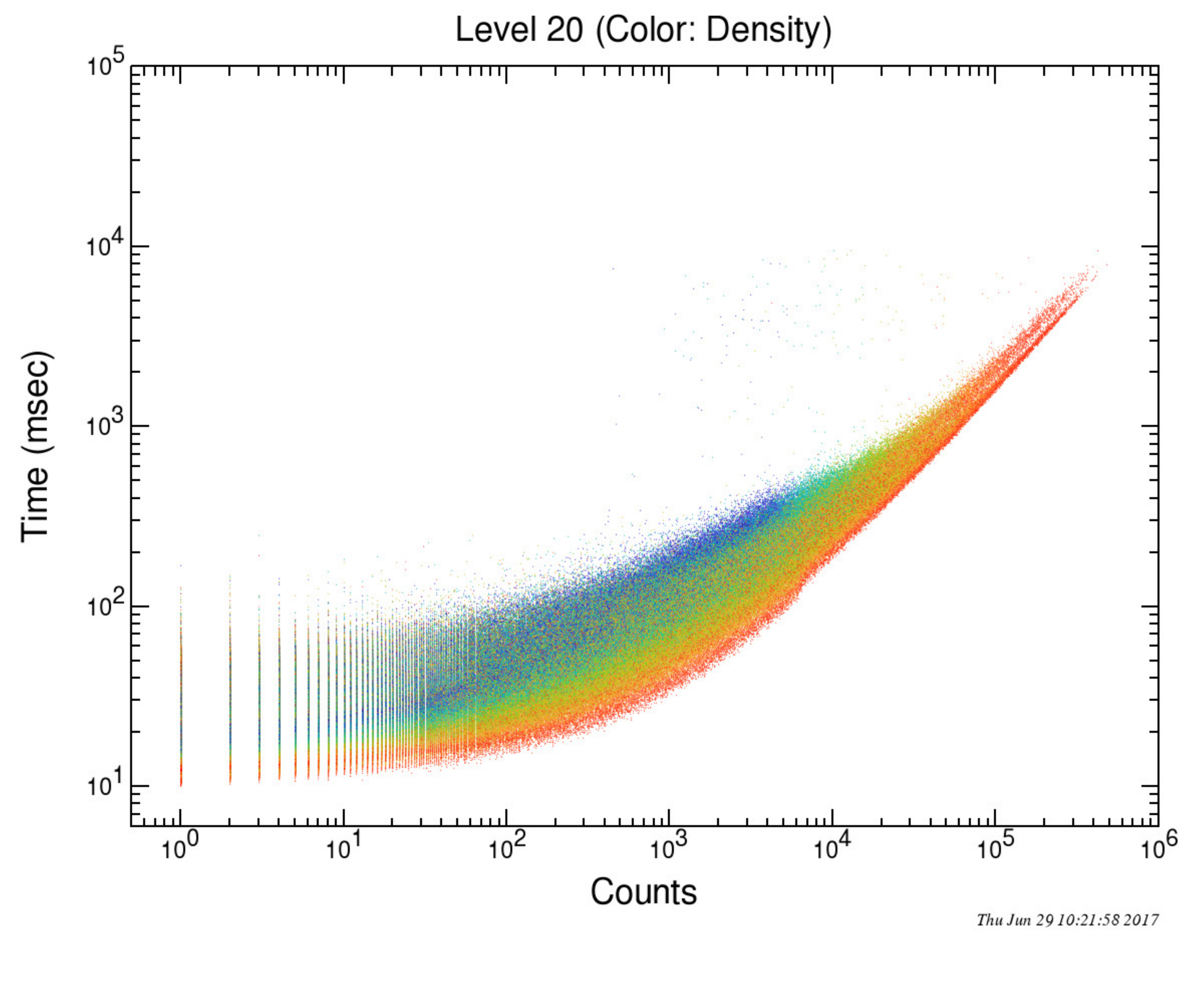}{Fig 2} {Distribution of query times versus number of records returned for the 2MASS catalog as the HPX indexing level under PostgreSQL increases from 7 to 20. The plots are color coded according to source density, with blue lowest and red highest.} 

\clearpage

\articlefiguretwo {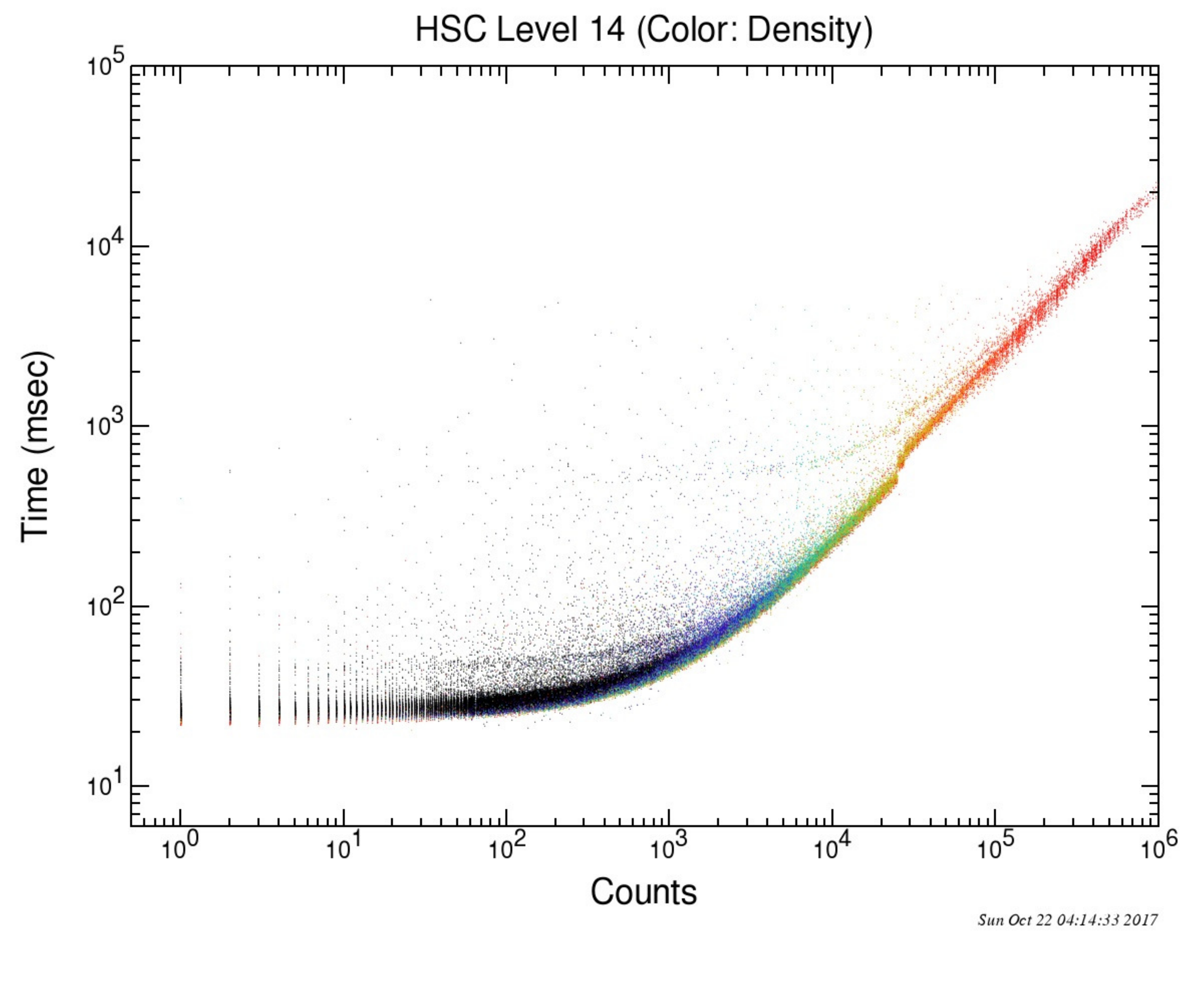}{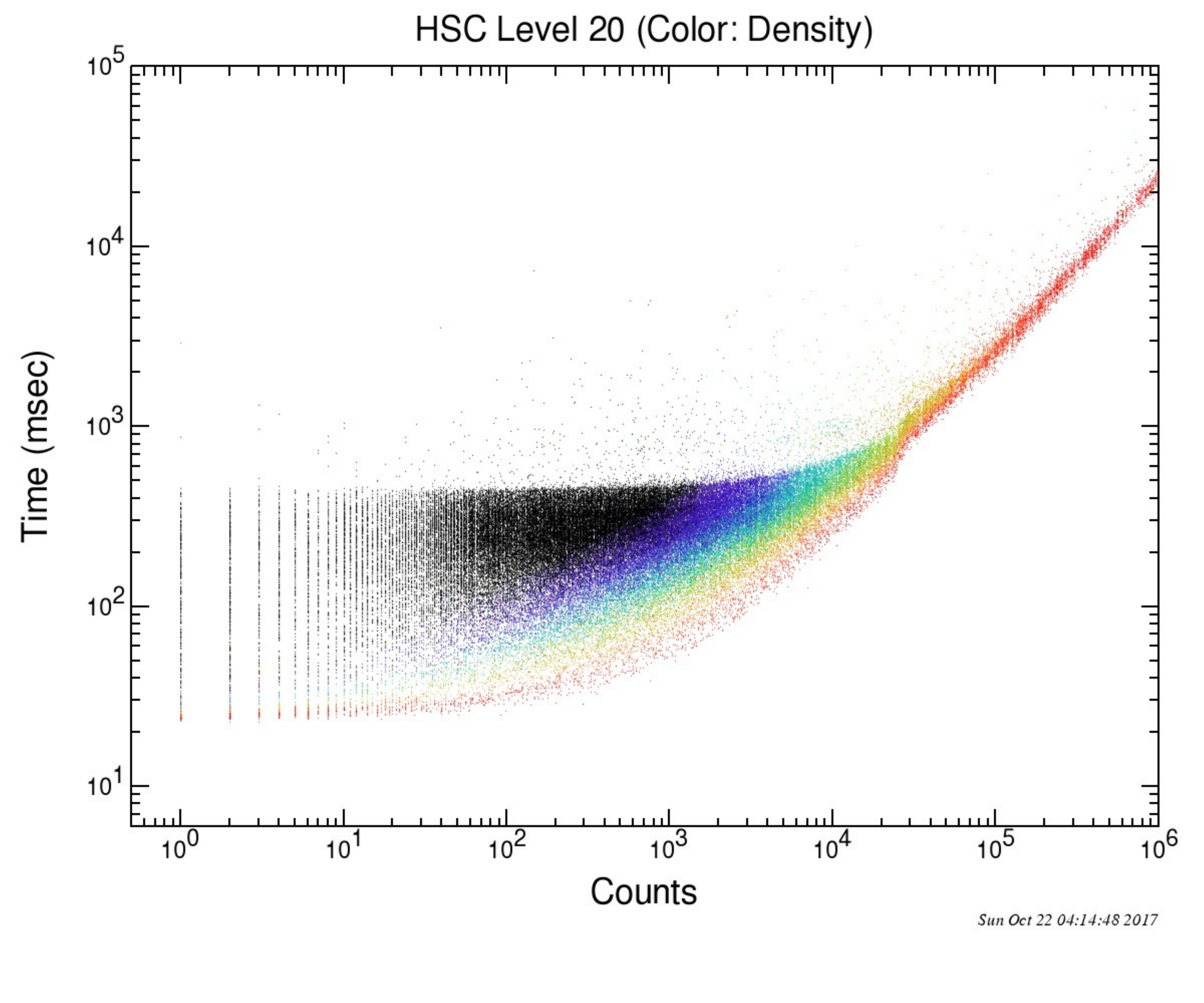}{Fig 3} {Distribution of query times versus number of records returned for the HSC for HPX indexing levels of 14 and 20 under PostgreSQL. The plots are color coded according to source density, with blue lowest and red highest. These plots contain timings for 100,000 points.} 

\articlefiguretwo {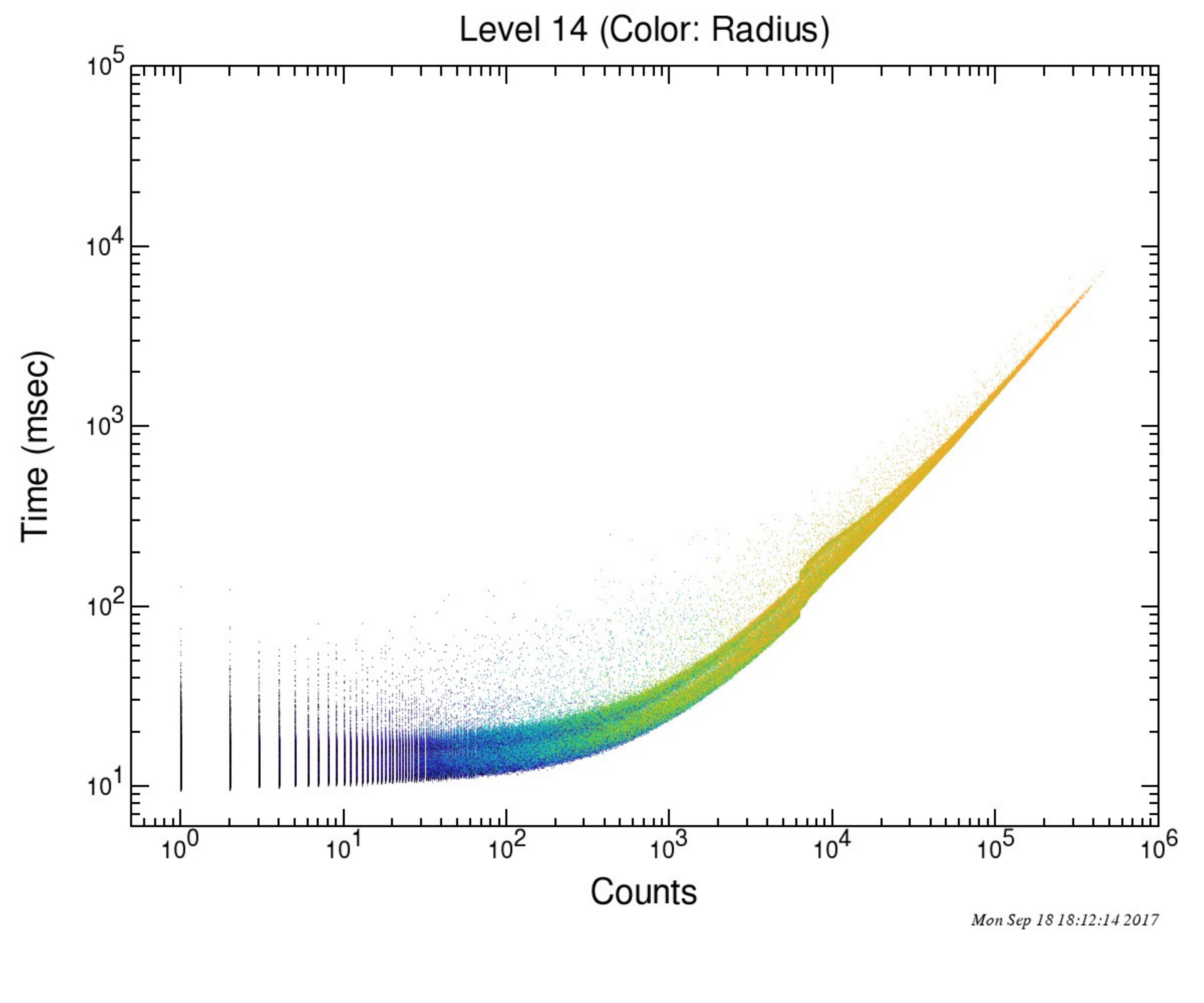}{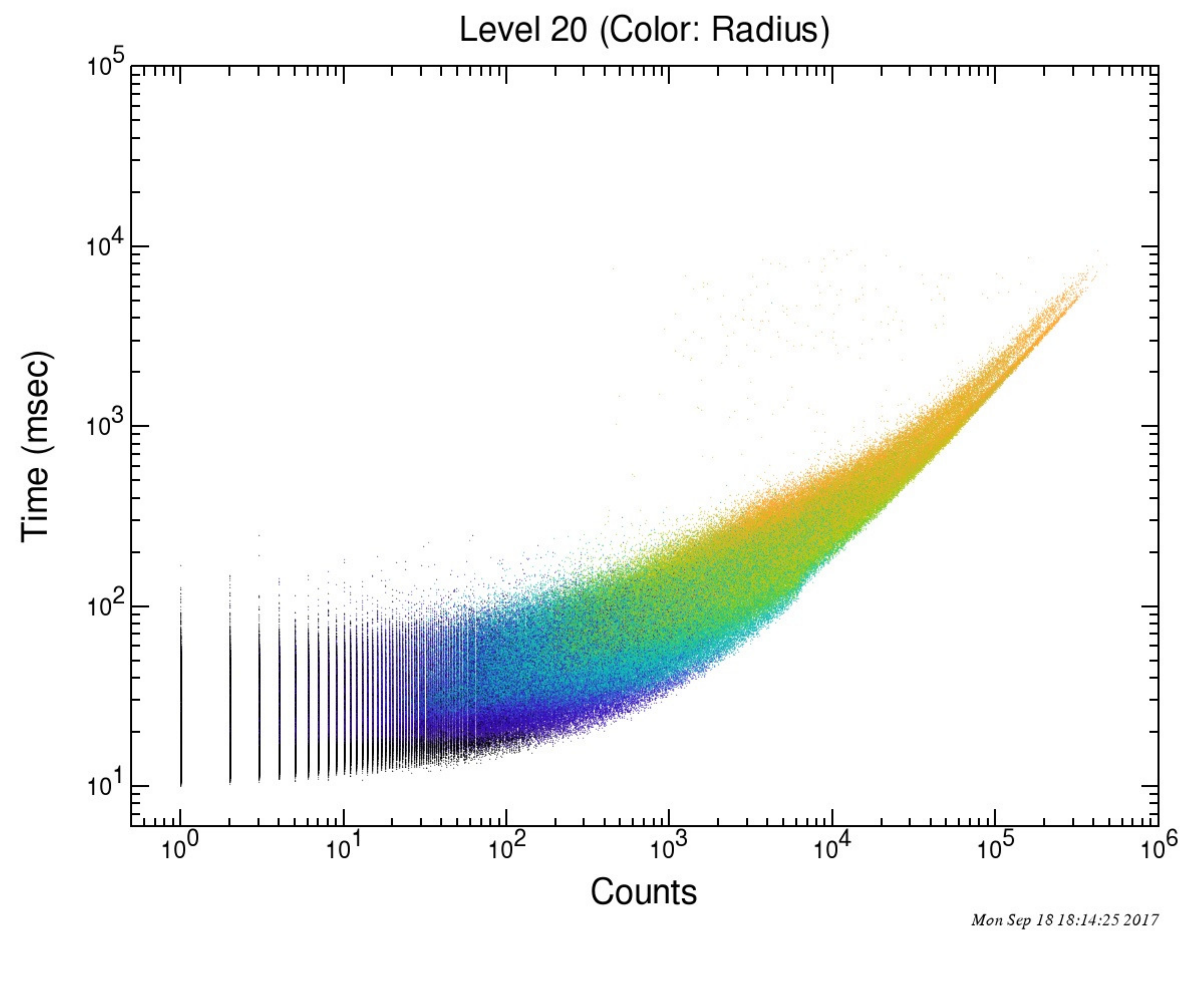}{Fig 4} {Distribution of query times versus number of records returned for the 2MASS Catalog for HPX indexing levels of 14 and 20 under PostgreSQL. The plots are color coded according to query radius, with blue lowest and red highest} 

\acknowledgements   Funding for the NASA Astronomical Virtual Observatories (NAVO)NAVO is provided by NASA through the Astrophysics Data Curation and Archival Research (ADCAR) program.  We thank Mr. Ricardo Ebert and Mr. Scott Terek for making the Solaris server available.

\bibliography{O4-6}  

\end{document}